\documentclass{article}
\usepackage{graphicx}
\usepackage{authblk}
\usepackage{subcaption}
\usepackage{amsmath}
\usepackage{amssymb}
\usepackage[a4paper, portrait, margin=1in]{geometry}

\usepackage[scaled]{helvet}

\usepackage[T1]{fontenc}

\newcommand{\Real}{\mathbb{R}}
\begin{document}

\title{A model for the generation of road networks}
\author[1,2,3,*]{Carlos Molinero}
\author[4]{Alberto Hernando}
\affil[1]{Austrian Institute of Technology}
\affil[2]{Complexity Science Hub, Vienna}
\affil[3]{Centre for Advanced Spatial Analysis, UCL}
\affil[4]{Kido Dynamics}
\affil[*]{\small{molinero@csh.ac.at}}

\maketitle
\begin{abstract}

As part of the effort undertaken to understand urban environments and their generation, we need to explore models that produce statistically valid configurations of roads. These sort of models will help us to derive plausible mechanisms for the spatial location of population.

This task is of fundamental importance, as we need to create an experimental environment that allows us to disentangle the specificities of a spatial configuration from the ideal system. Creating statistically valid models of road networks along with models of city generation allows to average the effects of geometry bringing us one step closer to understanding urban environments.

To completely understand road networks we need to be able to grasp what principles of economy do their growth entail. It is therefore of interest to explore the possible shape that a performance function would have for transportation systems.

In this work, we tackle this issue by proposing a network generation model based on a single parameter $\alpha$ which is capable of creating any type of network from trees to quasi-surfaces and which is shown to generate networks close to the real road networks under study. This is obtained through the definition of local and weighted versions of centrality measures. These centrality measures deal with distance-decay effects and nodes having different masses.

We set ourselves to determine the properties and different regimes of this $\alpha$-model and we lay out a definition for the performance of a network taking into account factors such as robustness, construction cost, congestion and distance. We obtain the optimal alpha from the analysis of the space of possible performance functions, giving an intuition on the self-organisational properties of the original network.

\end{abstract}

\section*{Introduction}

There are two main reasons for the existence of road networks. One is efficiency of travel in terms of effort, that is, an opened path is the least resistance option as opposed to moving through an uncategorised field. The second relates to specialisation of land uses, whereas the space needs to be partitioned into a differentiated movement strip to maintain accessibility of inner areas and the rest becomes functional space of some kind (residential, industrial, and so forth).

In this work we understand that specialisation of land use comes a-posteriori, once the least resistance path has been generated the left-over spaces acquire their function. This will not always be the case and for a given number of land uses with very restrictive needs, the roads will have to adapt to the remaining space, but this is to happen less often.

We can understand this \emph{effort} to generate a new path as a variation in the speed that I can travel at. The idea is similar to how paths are constructed in a field, the first person goes through the shortest path, breaking the vegetation in its path, making it easier for the second person to follow that path, go faster through it. Therefore, we then say that there are two speeds of travel, one through the road system and one outside of it and a person will not generate a new road as long as the time that it takes him/her to go through the existing roads is less than the time it would take to travel outside of the road system. We propose a framework for the generation of road networks based in this principle, and study their performance showing that there is an optimal configuration that resembles the actual road networks generated in real environments.


A few models of road generation have been proposed thus far, most of them define an iterative growth process of networks and roads. These models are supplied with new locations for population in a specific order which defines the final configuration of the network.

In some of this cases there is no global information, and the network is grown by local optimisation processes. In \cite{Barthelemy2008} the authors propose a local optimisation method that reproduces certain properties of networks such as the number of links vs. nodes. This is achieved by minimizing distances from the current road network to the new centres.  Other use indirect sources of global information, such as the work of \cite{Courtat2011} where a positive potential is defined around current developed areas that attract the creation of new locations in combination with a local repulsion potential to account for space being already occupied. Each new location subdivides the existing roads, thus forming the final road structure.

In other cases there exist a global performance function which have been defined in different manners along the literature. In \cite{bottinelli2017balancing} the authors propose an iterative growth approach that creates tree structures by choosing at each step between minimising total constructed distance by adding an edge to the network or maintenance costs by reconfiguring the network as the MST (minimum spanning tree). A similar function of global optimisation is proposed in \cite{Gastner2006a} where there is a trade-off between minimising total distances opposed to total length (which minimises maintenance costs). Both works create tree structures, in the case of \cite{Gastner2006a} the authors remark that real distribution networks show a remarkable optimisation level in both distances travelled and length of construction and propose a methodology to create these distribution networks.

The use of variations of the concept of distance \cite{Gastner2006optimal} is shown to be capable of creating a large range of structures. In that work the authors propose to minimise a distance function that is based in an equilibrium between metric distance and the topological distance in the path, tuned by the use of a parameter. This distance is then optimised globally by using a simulated annealing approach to find quasi-optimal solutions.

Far away from a modelling perspective, biological experiments with slime mould have shown that given the right input (food sources located where cities are) these creatures can reproduce networks that seem similar to man-made road networks \cite{strano2012vie,evangelidis2017physarum,zhang2015biologically}. The interest of these works lie in showing that if both networks show similar properties, then both must be self-organising and following similar optimisation strategies.

Other models, that are not intended for the creation of road networks have been proposed, some working with topological distances other focused on spatial networks and in most cases accounting with some definition of efficiency or optimal behaviour \cite{i2003optimization,brede2010coordinated,colizza2004network,barthelemy2006optimal,fabrikant2002heuristically}.

The concept of efficiency in networks has also been extensively treated in works such as \cite{latora2001efficient,vragovic2005efficiency,estrada2016communicability} and it is generally defined as the relation between distances travelled through the network and crow-flight distances.

The model proposed in this work is simultaneously capable of being used to obtain a road network given an input of cities and their populations, or to be used as an iterative growth model. We produce global information from the inputs by using a spatial interaction model, that calculates the number of trips between pairs of cities.

This road network model has a single parameter $\alpha$ that controls how dense will the network be, from trees to fully connected road networks. This $\alpha$ parameter controls the difference in speed between travelling along a road or through an open field. The networks produced are optimising the total distance travelled (as opposed to just the distance in the network) by taking into account the number of trips generated through the ordering of the pair of nodes for the construction of the roads. Simultaneously, through the use of the parameter $\alpha$ the total length of roads is also optimised. We then analyse the performance of each possible road network model produced by the different $\alpha$s and choose the one that is the optimal for the largest set of possible performance functions.


We lay out in this paper the basis for a measure of performance of road networks based on what we will call \textit{local and weighted centrality measures in networks}. Given a graph with different masses in its nodes (the population in our case) and considering that the transmission of information in the network has a distance decay we will calculate the number of trips between nodes and give a re-definition of closeness and betweenness. This will allow us to propose an ideal $\alpha$ value, that optimises the largest set from the space of possible performance functions. We also will define a measure of efficiency of a road network as the ratio between the performance of the real system and the performance of an idealised one, generated using our model with the optimal $\alpha$ as explained further on the text.

\section{The $\alpha$-model}
We propose a model for the generation of road networks that is based in a single parameter $\alpha$. The $\alpha$ parameter is the proportion between the speed of going through a road and the speed of going through the space without constructing a road. For example if we set $\alpha=.1$ and we consider that the speed limit of roads is $s=90 km/h$ then we are considering that the speed at which we can travel without a road is $s_0=9 km/h$. We use this parameter to consider how likely it is to construct a new road instead of using the roads that already exist.

A pre-requisite is to be able to calculate the number of trips ($N_{ij}$) to be generated for each pair of cities. This can be done with any of the existing spatial interaction models \cite{barbosa2018human,masucci2013gravity,yan2014universal,simini2012universal,stouffer1960intervening,wilson1971family}. We layout in the SI our own interpretation of this spatial interaction using the concept of {\it local and weighted} centrality measures.

\begin{figure}
\centering
\includegraphics[width=1\textwidth]{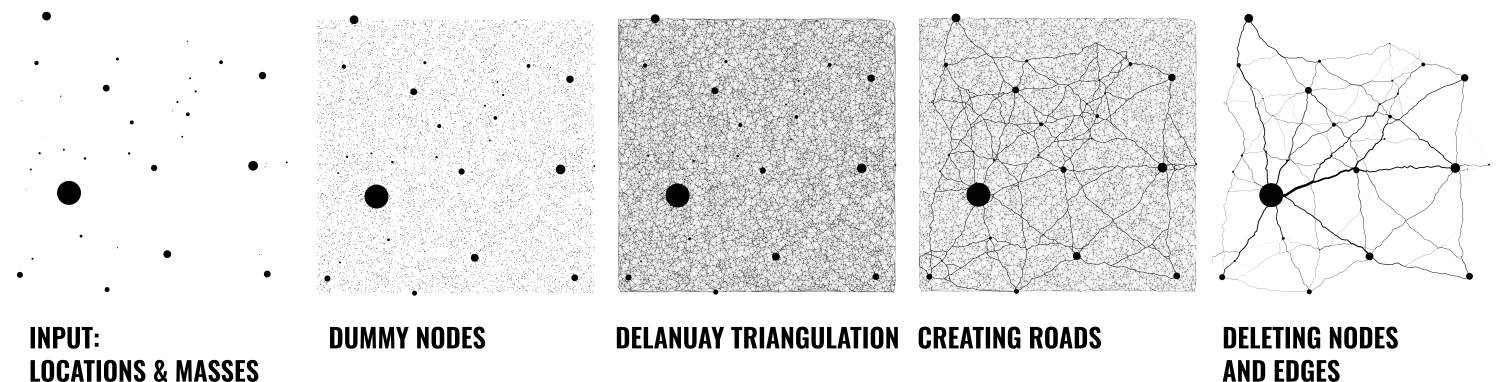}
\caption{Algorithm
\label{fig::algorithm}}
\end{figure}

\subsection{Algorithm}
We define a function $C: E\rightarrow \{0,1\}$ that returns whether an edge is a road. Given a parameter $\alpha\in (0,1]$, a set of locations and populations of cities and a distance decay function:
\begin{enumerate}
\item Add a vertex for each city with a mass equal to its population
\item We create a large number of dummy vertices in the space with mass equal $0$ (using a uniform random distribution) to serve as intermediate points. This serves to discretise the space.
\item We generate the delanuay triangulation of all the vertices in the space (the cities and the dummy nodes) we add all edges of the delanuay with $C=0$.
\item Order all the pairs of vertices by the number of trips that there is going to exist between them ($N_{ij}$), in decreasing order.
\item For each pair of vertices in decreasing order of number of trips:
\begin{enumerate}
\item calculate the shortest path in the delanuay from city A to city B. In order to calculate the shortest paths we set the weight of each edge to be $w(i)=\{\alpha l(i) | C(i)=1\}$ and $w(i)=\{l(i) | C(i)=0\}$ with $l(i)$ being the length of the segment.
\item tag every segment in the path to be "roads" (set $C=1$). Update the weights of the path using the new $C$ value.
\item continue with the next pairs of cities until all pairs are calculated. Then remove all edges of the graph with $C=0$ and all vertex with degree $k=0$.
\end{enumerate}
\end{enumerate}

\begin{figure}[t]
\center{\includegraphics[width=1\linewidth]{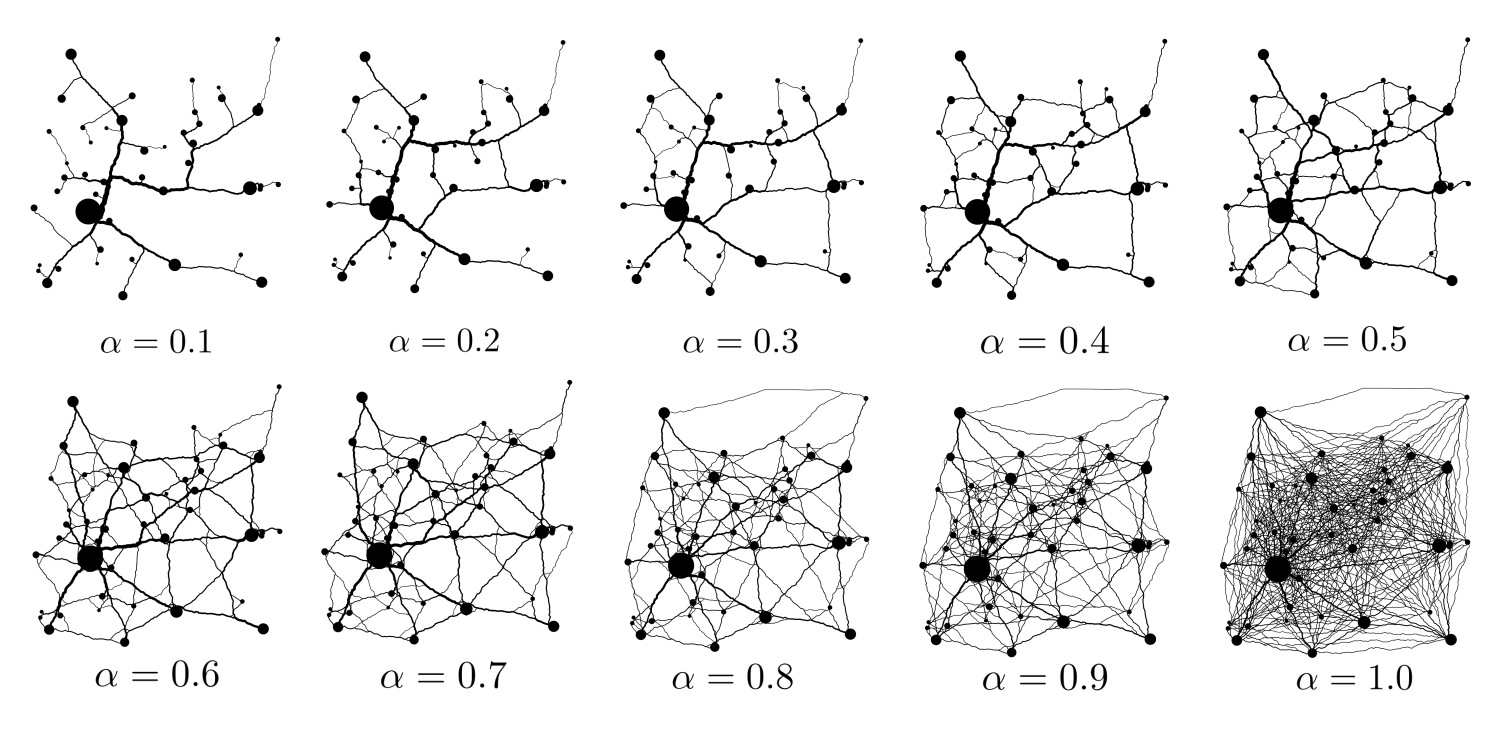}}
\caption{Roads generated with our $\alpha$-model for different values of $\alpha$.\label{fig::roadsBasedInAlpha}}
\end{figure}

In Fig.~\ref{fig::algorithm} we can see a mock up of the algorithm and in Fig.~\ref{fig::roadsBasedInAlpha} we can see the different results based in the $\alpha$ parameters.
This model generates a set of road networks that depending on its parameter $\alpha$ will be closer to trees (when $\alpha\rightarrow 0$) or roads traversing across the full space using one road for each pair of cities (when $\alpha=1$) passing through a phase where we encounter networks with typical connectivity profiles found in nature.

\section{Optimisation of $\alpha$}
From all the possible parameter space of the model (the different $\alpha$ values) there is only a handful of them that are optimal.

Very low values of $\alpha$ will generate trees, that are not robust and any failure on a link of the network will create disconnected cities and on top of it, all vehicles need to travel through the same links, thus, congestion will make the network non-optimal. On the other hand, for very large values of $\alpha$s we have networks that are connecting with one road every pair of cities, although these structures will have attractive characteristics such as minimizing the trip distance between the cities and their high resilience, they generate a large number of intersections that in turn produce congestion and reduce the optimality of the network configuration. On top of it we need to account for the cost of construction of a road network which will be elevated in this last case since it will be proportional to the total length  of the network.

Therefore, it stems that intermediate values of $\alpha$ will be the ones with a highest fitness reaching an optimal configuration when balancing the factors of robustness, distances, cost of construction of the network and congestion factors.

To conduct the experiments, we averaged our results over 1500 repetitions for each possible value of $\alpha$. In each of these repetitions, the locations of the vertices that represent the cities are randomised, in order to obtain results that are exempt of particularities in the spatial distribution of cities.

We define the formula of our performance function as
\begin{equation}
\label{eq::performance}
\Phi=\frac{r}{l}\left(\frac{1}{d}\right)^{\beta_1}\left(\frac{1}{vn}\right)^{\beta_2}
\end{equation}

where we can see that the factors that affect performance are separated into 3 categories.
\begin{enumerate}
\item The first factor is ${r}/{l}$ where $r$ stands for robustness and $l$ for the total length of the network. We consider this factor to be a fundamental function of performance of a network, one that all networks must use at some extent. Robustness is the fundamental reason why to construct a network in the first place otherwise trees are cheaper to build. Robustness represents the duplicities of connections between vertices of the graph. The total length of a network is proportional to the cost of its construction, larger networks will require more energy and resources to build and networks will try to minimise this quantity.
\item The second factor $\left({d}\right)^{-\beta_1}$ represents the inverse of the total distance that is traveled between vertices of the graph. This is an important property of networks, but there could be the case of networks of different classes not needing to optimise this factor. We set a exponent $\beta_1\in [0,1]$ to weight the effect of this factor.
\item The third factor $\left({vn}\right)^{-\beta_2}$ relates to optimisation factors that relate to congestion. The quantity $v$ is the total volume of flow that is traversed on the generated trips which is proportional to the congestion generated through traffic. Finally $n$ is the number of intersections that are traversed in the total number of trips which will also generate congestion through the waiting times.
 We set a exponent $\beta_2\in [0,1]$ to weight the effect of this factor.
\end{enumerate}

The specific formulae for each factor of the performance function can be found in the SI of this paper.



\begin{figure}[t]

\center{\includegraphics[width=1\textwidth]{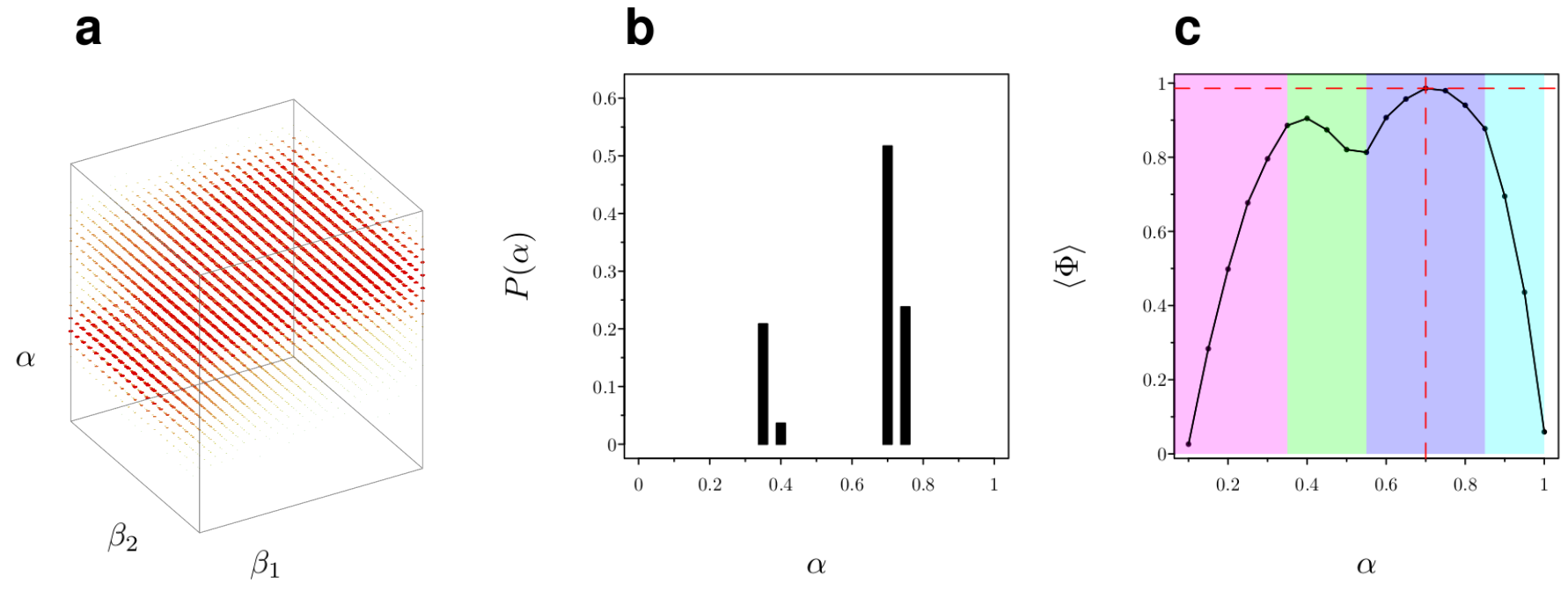}}
\caption{Result of measuring performance in the parameter space of $\beta_1$ and $\beta_2$. a) Optimal $\alpha$ for each possible value, showing that the results are discrete and that for most values of the parameter space there is a single $\alpha$ that optimises them. b) Histogram of the optimal $\alpha$, showing that in most cases our performance function is optimised for $\alpha=0.7$. c) Performance function averaged over all possible $\beta_1$ and $\beta_2$, showing that the best model is around $\alpha=0.7$.  \label{fig::parameterSpace}}
\end{figure}

\begin{figure}
\centering
\includegraphics[width=1\textwidth]{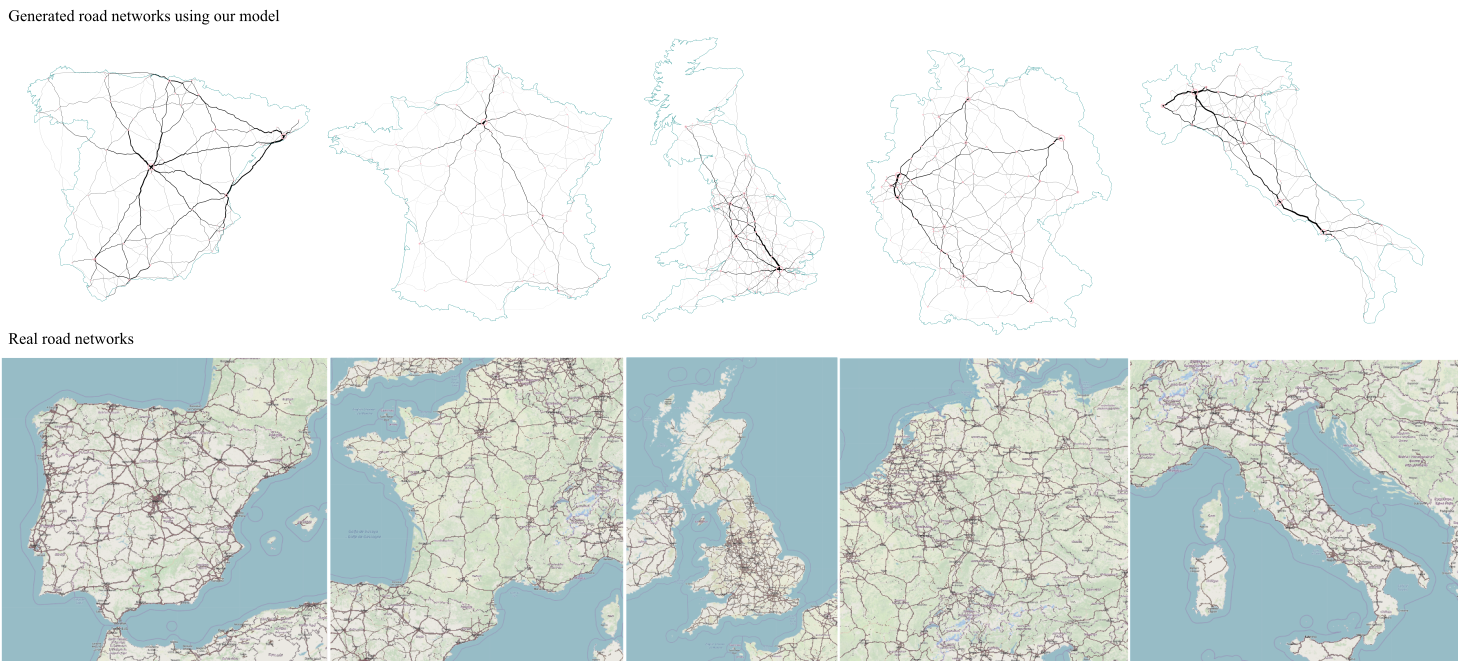}
\caption{Comparison between generated road networks with the optimal $\alpha=0.7$ value obtained and real road networks. We can observe that both model and roads follow a similar distribution. This can be improved by introducing information in the model about the topography of the country (which is the source of most divergences between the result of the model and reality) as shown in the SI.
\label{fig::countries}}
\end{figure}

We can imagine that different classes of networks will optimise different proportions of these 3 factors. We have therefore explored the parameter landscape for combinations of $(\beta_1,\beta_2)$ that produce different performance functions and extracted which was the optimal $\alpha$ for each case, obtained at the maximum of Eq.~\ref{eq::performance}.

%

This parameter space returns optimal $\alpha$s in discrete steps $\alpha \in$ \{0.35,0.4,0.7,0.75\} as can be seen if we observe the histogram of the parameter space (Fig.\ref{fig::parameterSpace}b) we can see that there are more combinations of exponents $(\beta_1,\beta_2)$ that give an optimal $\alpha=0.7$. We can also observe in Fig.\ref{fig::parameterSpace}c that when calculated the average performance for any possible performance function, the values that behave better are around $\alpha=0.7$, we believe that using this range of values as our optimal parameter ensures that for any possible performance function the generated networks are going to be generated efficiently.

 In order to appreciate the quality of the road networks generated, we show in Figure~\ref{fig::countries} the networks produced in several European countries, and how closely they resemble their real counterparts. These road networks were generated with partial information (the only input were cities above $5\times 10^4$ people). By including other information such as topography, river networks, etc. we could increase the quality of the final road network (see SI) given that topography is the largest source of divergence between the model and reality.

\section{Network measures and $\alpha$ of the real road network}
Similarly to previous works on road network generation, we will show a few measures that characterise the resulting graphs using our model. This will serve to indirectly obtain the $\alpha$ value that would create the closest network to the existing one and therefore infer its $\alpha$ parameter.

In order to have comparable results, we first need to discard from the real road network every road that is not used to carry traffic from our set of origin/destinations (the cities the model uses as inputs), otherwise measurements are not comparable. The simplest way to achieve this is to set every node's mass to 0 and then choose one node per city and assign the total mass of the city, then calculate the flows (local-weighted betweenness) and sub-select only roads that have some flow going through them. We should have the speed limits of each road to calculate the time distance, otherwise the skeleton would not correspond to reality. In our case we do not have those speed limits, and as a substitute we calculate the hierarchical index of the angular percolation \cite{Molinero2017angular} because it relates to the classification of roads, which will be proportional to the speeds and we set our speeds to $I_i+1$ before we calculate the flows.

\begin{figure}
\centering
\includegraphics[width=1\textwidth]{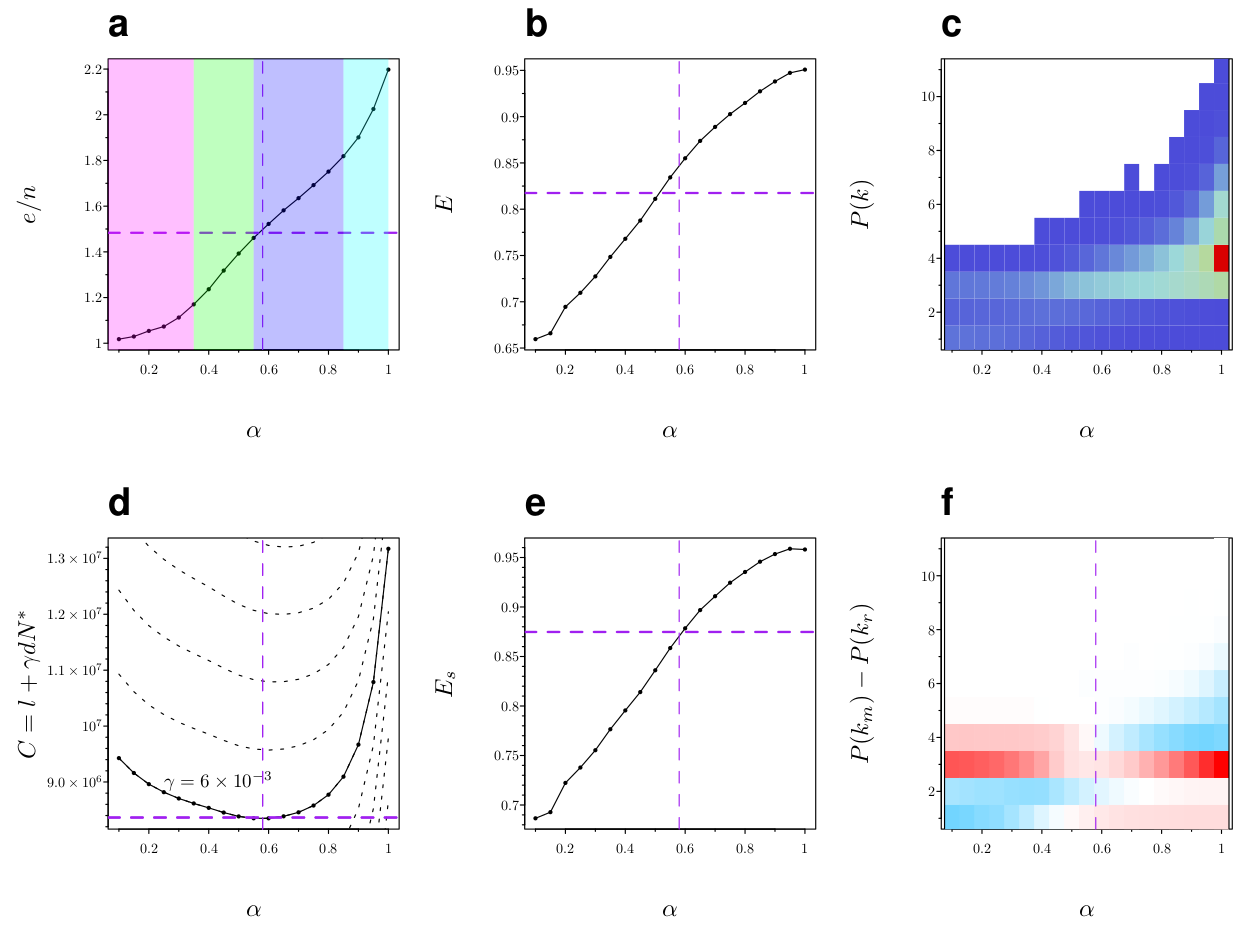}
\caption{Network measurements. Horizontal purple line (colour online) corresponds to the measurement taken for the real road network. Vertical purple line is $\alpha=0.58$, which is the value that seems to be a better fit with the real road network measurement. Vertical grey lines correspond to the $\alpha$ values calculated as optimal using the performance function proposed. a) ratio of edges to nodes \cite{Barthelemy2008}; b) Efficiency as the normalised ratio between euclidean distance and network distance \cite{latora2001efficient}; c) distributions of degrees of nodes for the model; d) performance function as defined in \cite{Gastner2006optimal}; e) Efficiency, in this case calculated on the structural graph; f) difference in the distribution of degrees between the real road network and the model.
\label{fig::networkMeasurements}}
\end{figure}

As we can observe in Fig.~\ref{fig::networkMeasurements}.a, the ratio of edges against nodes of the real road network seems to indicate that the road network would correspond to the model produced by an $\alpha=0.58$. Instead the efficiency measured directly in the network as is (Fig.~\ref{fig::networkMeasurements}.b) is lower than in the model, this is because the model does not have to avoid any obstacles (airports, rivers, agricultural fields) while the real road network does, which produces winding roads. In order to correct for this, and try to have comparable systems, we produce the structural graph (nodes of degree 2 are removed and weights are recalculated producing straight lines between intersections) for both the real network and the model and measure its efficiency in the same manner (Fig.~\ref{fig::networkMeasurements}.e) to remove the effect of artefacts produced by avoiding obstacles. This last measurement shows that both networks then coincide in the same approximate $\alpha$. This is also the case for the difference in the distribution of degrees (Fig.~\ref{fig::networkMeasurements}.f) where white values correspond to 0 difference (red is negative and blue positive) showing that again both models coincide in the same $\alpha$. Moreover, by looking at the efficiency (Fig.~\ref{fig::networkMeasurements}.e) we can observe that there is a linear increase with respect to the parameter $\alpha$ up to around $\alpha=[0.65,0.7]$ hinting that networks above this values behave differently and are no longer optimal.

We see in Fig.~\ref{fig::networkMeasurements}.d the performance function defined in \cite{Gastner2006optimal}. This performance is defined as the cost of the network, understood as the linear combination of the total length of the network and the distance travelled in the system. The authors use a simplified assumption on the number of trips between locations devoid of a distance decay. In our case we have adjusted it to our framework as $C=l+\gamma dN^*$, where $\gamma$ has the same meaning as the one in the original paper. This parameter weights the combination of the two factors and by modifying it, the optimal $\alpha$ could be any (when $\gamma=0$ only length is taken into account and therefore a tree is better and when it goes to infinity only distances of trips are taken into account and a surface is better). We show several possible performance functions for different $\gamma$ in dotted points, while we have drawn as a line the one that corresponds to $\gamma=6\times 10^-3$, which has an optimal value in $\alpha=0.6$. We have done so, because it is the only value of $\gamma$ for which the measurement of the real road network, coincides with the value of the model at the optimal point. This, again, hints to $\alpha=0.58$ as the actual value of the road network.

\begin{figure}
\centering
\includegraphics[width=1\textwidth]{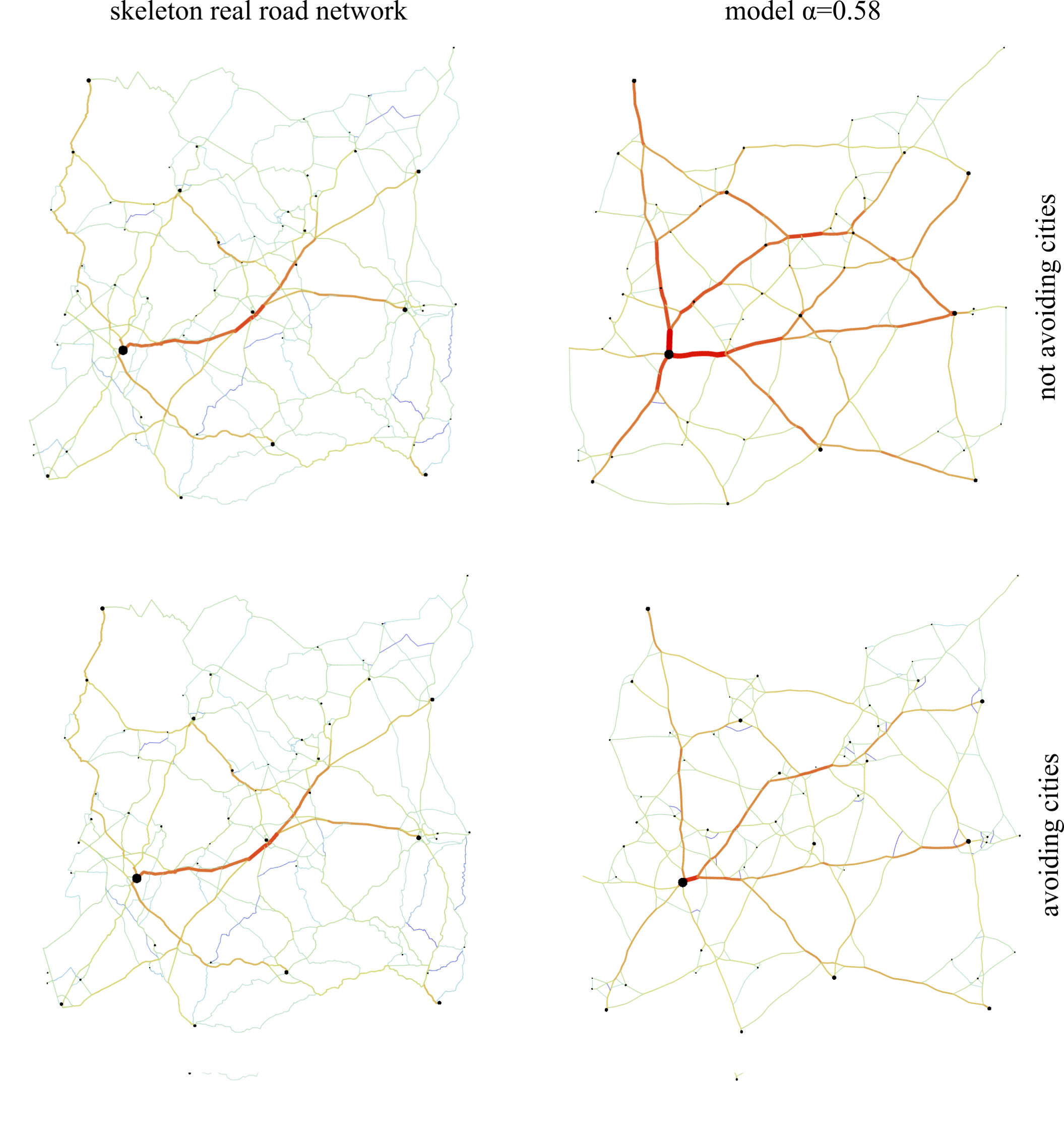}
\caption{Comparison between the skeleton of the road network and the model with the $\alpha=0.58$ obtained from the measurements. In the upper row, the model is the simplest possible model, where all speeds in the delanuay are equal. The lower row is the model produced by avoiding city centers (this is further explained in the SI).
\label{fig::comparisonWithRealNetwork}}
\end{figure}

We can therefore expect that $\alpha=0.58$ is the value that would produce a closer model to the real road network, even though it is slightly below the optimal values measured using our performance function. We attempt to show the comparison between a real road network and our model using this value of $\alpha$ in Fig.\ref{fig::comparisonWithRealNetwork}.

Why the $\alpha=0.58$ of the real road network diverges from the expected value of $0.65-0.7$  could be due to an incorrect definition of the performance function or to problems in the implementation and design of road networks. One reason could be the measurement of congestion ($v$) which in reality is not problematic until the capacity of the lanes is surpassed, but we assumed a value that is proportional to any flow because no information on the capacity of lanes exists. Another reason could be that the optimal values of $\alpha$s do not really exist in reality, because other constrains interfere with their design leading to close to optimal but never optimal results. A third, and very probable explanation is that $\alpha$ might depend on the number of trips. This means that connections between large cities (main highway system) would have an $\alpha$ close to 0.7 and as the number of trips decreases $\alpha$ would drop down to around 0.6 in order not to produce new roads for just a few trips.

We can further observe that even the simplest measure (Fig.~\ref{fig::networkMeasurements}.a) already shows 4 different slopes indicating the existence of 4 different regimes in the model: $0<\alpha\leq 0.35$, which corresponds to quasi-trees and trees. $0.35<\alpha\leq 0.55$ for networks that are closer to trees, $0.55<\alpha\leq 0.85$ which is the realm of networks and $0.85<\alpha\leq 1$ that grow fast into surface-like systems. This different regimes are also visible in our performance function as shown in Fig.~\ref{fig::parameterSpace}c.

\section{Conclusions}

We have presented in this work an approach to generate realistic road networks to serve as a basis for the experimentation in the field of urban science. The model provides a very simple algorithm and a behaviour characterised basically by a single parameter. That is not to say that other factors are not involved in the creation of the road network. Different spatial interaction models would modify the ordering in which roads are constructed, and therefore provide different results. This is actually an interesting property, since we expect that the correct formulation of a spatial interaction model combined with the correct $\alpha$ should basically produce a replica of the real road network, and therefore it could be used as testing grounds for different models of spatial interaction.

We have presented in the SI the concept of local-weighted centrality measures. Furthermore, coming from the formulation of the number of trips, we propose to use $N_{ij}$ as an ordering function to a model that creates a hierarchical planar and spatial network based on the parameter $\alpha$. This $\alpha$ controls how spread or concentrated is the resulting network.

We also layout the basis for a performance function for spatial networks that takes into account three types of factors, a fundamental one of robustness against length of network to construct, a second factor that minimises the distance travelled in the network per trip and a third one that relates to congestion factors (the volume of flow and number of intersections traversed on trips). We study the parameters that maximises each possible performance function, and choose to keep the one that is statistically more frequent.

Our measurements seems to indicate the existence of 4 types of regimes in the model distinguishing between trees, networks half way between trees and true networks, networks and quasi-surfaces.


\bibliographystyle{plain}

\begin{thebibliography}{10}

\bibitem{barbosa2018human}
Hugo Barbosa, Marc Barthelemy, Gourab Ghoshal, Charlotte~R James, Maxime
  Lenormand, Thomas Louail, Ronaldo Menezes, Jos{\'e}~J Ramasco, Filippo
  Simini, and Marcello Tomasini.
\newblock Human mobility: Models and applications.
\newblock {\em Physics Reports}, 734:1--74, 2018.

\bibitem{barthelemy2006optimal}
Marc Barth{\'e}lemy and Alessandro Flammini.
\newblock Optimal traffic networks.
\newblock {\em Journal of Statistical Mechanics: Theory and Experiment},
  2006(07):L07002, 2006.

\bibitem{Barthelemy2008}
Marc Barth\'{e}lemy and Alessandro Flammini.
\newblock {Modeling Urban Street Patterns}.
\newblock {\em Physical Review Letters}, 100(13):138702, April 2008.

\bibitem{bottinelli2017balancing}
Arianna Bottinelli, R{\'e}mi Louf, and Marco Gherardi.
\newblock Balancing building and maintenance costs in growing transport
  networks.
\newblock {\em Physical Review E}, 96(3):032316, 2017.

\bibitem{brede2010coordinated}
Markus Brede.
\newblock Coordinated and uncoordinated optimization of networks.
\newblock {\em Physical Review E}, 81(6):066104, 2010.

\bibitem{colizza2004network}
Vittoria Colizza, Jayanth~R Banavar, Amos Maritan, and Andrea Rinaldo.
\newblock Network structures from selection principles.
\newblock {\em Physical review letters}, 92(19):198701, 2004.

\bibitem{Courtat2011}
T~Courtat, C~Gloaguen, and S~Douady.
\newblock {Mathematics and Morphogenesis of the City: A Geometrical Approach}.
\newblock {\em Physical Review E}, pages 0--13, 2011.

\bibitem{estrada2016communicability}
Ernesto Estrada and Naomichi Hatano.
\newblock Communicability angle and the spatial efficiency of networks.
\newblock {\em SIAM Review}, 58(4):692--715, 2016.

\bibitem{evangelidis2017physarum}
Vasilis Evangelidis, Jeff Jones, Nikolaos Dourvas, Michail-Antisthenis
  Tsompanas, Georgios~Ch Sirakoulis, and Andrew Adamatzky.
\newblock Physarum machines imitating a roman road network: the 3d approach.
\newblock {\em Scientific reports}, 7(1):7010, 2017.

\bibitem{fabrikant2002heuristically}
Alex Fabrikant, Elias Koutsoupias, and Christos~H Papadimitriou.
\newblock Heuristically optimized trade-offs: A new paradigm for power laws in
  the internet.
\newblock In {\em International Colloquium on Automata, Languages, and
  Programming}, pages 110--122. Springer, 2002.

\bibitem{freeman1978centrality}
Linton~C Freeman.
\newblock Centrality in social networks conceptual clarification.
\newblock {\em Social networks}, 1(3):215--239, 1978.

\bibitem{Gastner2006a}
Michael~T Gastner and M~E~J Newman.
\newblock {Shape and efficiency in spatial distribution networks}.
\newblock {\em Journal of Statistical Mechanics: Theory and Experiment},
  2006(01):P01015--P01015, jan 2006.

\bibitem{Gastner2006optimal}
Michael~T Gastner and MEJ Newman.
\newblock Optimal design of spatial distribution networks.
\newblock {\em Physical Review E}, 74(1):016117, 2006.

\bibitem{hernando2014space}
Alberto Hernando, R~Hernando, and Alberto Plastino.
\newblock Space--time correlations in urban sprawl.
\newblock {\em Journal of The Royal Society Interface}, 11(91):20130930, 2014.

\bibitem{i2003optimization}
Ramon~Ferrer i~Cancho and Ricard~V Sol{\'e}.
\newblock Optimization in complex networks.
\newblock In {\em Statistical mechanics of complex networks}, pages 114--126.
  Springer, 2003.

\bibitem{latora2001efficient}
Vito Latora and Massimo Marchiori.
\newblock Efficient behavior of small-world networks.
\newblock {\em Physical review letters}, 87(19):198701, 2001.

\bibitem{masucci2013gravity}
A~Paolo Masucci, Joan Serras, Anders Johansson, and Michael Batty.
\newblock Gravity versus radiation models: On the importance of scale and
  heterogeneity in commuting flows.
\newblock {\em Physical Review E}, 88(2):022812, 2013.

\bibitem{Molinero2017angular}
Carlos Molinero, Roberto Murcio, and Elsa Arcaute.
\newblock The angular nature of road networks.
\newblock {\em Scientific reports}, 7(1):4312, 2017.

\bibitem{sabidussi1966}
Gert Sabidussi.
\newblock {The centrality index of a graph}.
\newblock {\em Psychometrika}, 31(4):581--603, December 1966.

\bibitem{simini2012universal}
Filippo Simini, Marta~C Gonz{\'a}lez, Amos Maritan, and Albert-L{\'a}szl{\'o}
  Barab{\'a}si.
\newblock A universal model for mobility and migration patterns.
\newblock {\em Nature}, 484(7392):96, 2012.

\bibitem{stouffer1960intervening}
Samuel~A Stouffer.
\newblock Intervening opportunities and competing migrants.
\newblock {\em Journal of regional science}, 2(1):1--26, 1960.

\bibitem{strano2012vie}
Emanuele Strano, Andrew Adamatzky, and Jeff Jones.
\newblock Vie physarale: Evaluation of roman roads with slime mould.
\newblock {\em arXiv preprint arXiv:1204.2260}, 2012.

\bibitem{vragovic2005efficiency}
I~Vragovi{\'c}, E~Louis, and Albert D{\'\i}az-Guilera.
\newblock Efficiency of informational transfer in regular and complex networks.
\newblock {\em Physical Review E}, 71(3):036122, 2005.

\bibitem{wilson1971family}
Alan~Geoffrey Wilson.
\newblock A family of spatial interaction models, and associated developments.
\newblock {\em Environment and Planning A}, 3(1):1--32, 1971.

\bibitem{yan2014universal}
Xiao-Yong Yan, Chen Zhao, Ying Fan, Zengru Di, and Wen-Xu Wang.
\newblock Universal predictability of mobility patterns in cities.
\newblock {\em Journal of The Royal Society Interface}, 11(100):20140834, 2014.

\bibitem{zhang2015biologically}
Xiaoge Zhang, Andrew Adamatzky, Felix~TS Chan, Yong Deng, Hai Yang, Xin-She
  Yang, Michail-Antisthenis~I Tsompanas, Georgios~Ch Sirakoulis, and Sankaran
  Mahadevan.
\newblock A biologically inspired network design model.
\newblock {\em Scientific reports}, 5:10794, 2015.

\end{thebibliography}

\newpage
\renewcommand{\thesection}{S\arabic{section}}
\renewcommand{\thefigure}{S\arabic{figure}}
\setcounter{figure}{0}
\setcounter{section}{0}

\section*{Supplementary information}

\section{Local-weighted centrality measures in networks}
\label{sec::centrality}

Centrality measures in networks originated in the field of sociology with \cite{sabidussi1966,freeman1978centrality}. A vertex represented a person, and an edge a friendship. Those edges were usually not weighted, for example in the spread of a rumor, the rumor spreads discretely, one step at a time, with each conversation and to calculate who would be the most important person in the network, it was assumed that the rumor reached everyone, and just once regardless of the distance.

Here we are working with road networks between cities. Cities have a wide population distribution, larger cities generating most of the traffic, while smallest cities create less drivers and an intersection does not generate any incoming or outcoming traffic at all. On top of it, as the distances increases from the origin city to the rest of the network, the number of trips will decay with a certain function of the distance. Therefore, we propose a modification to centrality measures for graphs with vertex of different masses that use a distance decay function accounting for the probabilities of making a trip at a certain distance.

We consider that a road network system is a graph $G=(V,E,w,m)$ where $V$ is a set of vertices, $E$ is a set of edges, $w:E\rightarrow \Real$ is a weight function of the edges and $m: V\rightarrow \Real$ is a function that returns the mass of a vertex.

In order to calculate how many trips will be created between two different cities we need to establish a distance-decay function that determines how likely a trip is going to be depending exclusively on the distance and mode of transport (in this case we use a single mode of transport). The choice of a Cauchy-Lorentz distribution for this probability is based on the correlations observed between the growth of cities \cite{hernando2014space} and has the following probability density function:
\begin{equation}
p_d(d_{ij})=\frac{2}{\pi d_0}\frac{1}{1+\left(\frac{d_{ij}}{d_0}\right)^2}
\end{equation}
where $d_{ij}$ is the distance from vertex $i$ to $j$, $d_0$ is a constant that weights the distance effect. $\frac{2}{\pi d_0}$ is just the normalisation factor for a variable $d$ with domain $[0,\infty)$ which gives us the probability of a trip happening at a certain distance. As a generalisation, for other types of networks with different phenomena happening upon it, this distance-decay function could take any other distribution and as such can be considered a parameter to local-weighted centrality measures.

The probability of making a trip from one vertex to another vertex will be a function of how much attractiveness the vertex has (that we will consider proportional to its mass) and the distance decay, therefore the probability of making an outgoing trip from vertex $i$ to vertex $j$ is
\begin{equation}
p_o(i,j)=\frac{p_d(d_{ij})m(j)}{\sum_{\forall k \in V} p_d(d_{ik})m(k)}
\end{equation}
where $m(j)$ is the mass of vertex $j$. Note that the normalisation factor $\sum_{\forall k \in V}p_d(d_{ik})m(k)$, include all internal trips from $i\rightarrow i$. Now we can finally asses the final number of trips from vertex $i$ to $j$ by including the mass in $i$ as
\begin{equation}
\label{eq::trips}
N_{ij}=p_o(i,j)m(i)
\end{equation}

Let the reader notice that if we expand equation \ref{eq::trips} we recover a gravity model such as the ones typically used in spatial interaction models.
\begin{equation}
N_{ij}=C\frac{m(i)m(j)}{1+\left(\frac{d_{ij}}{d_0}\right)^2}
\end{equation}
where $C$ represents a normalisation factor.

We can now define local centrality measures using the distribution of trips within the network. As such, we say that the local-weighted betweenness of a segment $k$ is
\begin{equation}
\label{lwbetweenness}
b(k)=\sum_{\forall i}\sum_{\forall j}\frac{N_{ij}(k)}{\sigma_{ij}}
\end{equation}
where $N_{ij}(k)$ is an abuse of notation for the function that returns $N_{ij}$ when $k$ is in the shortest path from $i$ to $j$ and 0 otherwise. $\sigma_{ij}$ is the number of shortest paths from $i$ to $j$ (which in real-weighted networks is usually 1). Note that because intersections in the middle of the road have a mass of 0 they will generate no trips ($N_{ij}=0$) so we can safely apply this measure to the whole network without any further consideration.

We can also consider a measure of local-weighted closeness for a node $i$:
\begin{equation}
c(i)=\frac{1}{\sum_{\forall j} N_{ij}d_{ij}}
\end{equation}

Note that if we set a uniform probability as our distance decay (no distance decay) and all masses are $1$ we recover the typical centrality measures. This means that centrality measures can be understood as a special case of our formulation.




\section{Calculating the performance function}
Robustness is a fundamental property of many types of networks. It assures that the network is resilient against breakdowns or attacks and that the network structure has enough duplicities such that all nodes will remain reachable. In most of the research literature robustness is measured as the number of links to be removed before generating 2 components, considering that each edge has the same probability of being removed and counting the fraction of the number of edges removed before breaking up the network. This is again difficult to justify in the case of road networks, most of the factors that will end up in an edge of the network being disconnected depend linearly on the length of the link and not upon its mere existence. That is the case for phenomena that spread over the space (such as snow, rain or floods) and failure of the asphalt material cracking the road. Therefore, in order to measure this property we consider that the probability of a meter of road getting damaged is a certain constant value $\beta$ (since it depends on the properties of the material/space not on the network configuration) and then we remove pieces of 1 meter from the system following this probability (pieces which may or may not belong to the same link of the network). We measure how many meters could be removed from the network before having a city (vertex of mass larger than 0) disconnected from the rest. Since the networks have different lengths, this value ($r$) is normalised using the total length of the road network making that our indicator for robustness is the fraction of meters that needs to fail from the network before disconnecting a city.

The total cost of construction and maintenance of a road network will be proportional to its total length, so we define $l=\sum_{i\in E}{w(i)}$ as the aggregation of the lengths of all edges in the network and expect that this value is minimised in the resulting network.

In order to measure the performance we need 3 other factors: the average distance traveled by all vehicles in the system
\begin{equation}
d=\frac{1}{N^*_{ij}}\sum_{\forall i}\sum_{\forall j} N_{ij}d_{ij}
\end{equation}
the volume of flow traversed in their trip. Which is measured as how many meters need to be travelled under large flows of cars in all the trips produced in the system.

\begin{equation}
v=\frac{1}{N^*_{ij}}\sum_{\forall i}\sum_{\forall j}\sum_{k\in \overrightarrow{ij}} \frac{b(k)}{N^*_{ij}}w(k)N_{ij}(k)
\end{equation}
where $N^*_{ij}=\sum_{\forall i}\sum_{\forall j\neq i}N_{i,j}$ is the number of outgoing trips in the system that serves to normalise the values for different networks, $b(k)$ is the local-weighted betweenness of segment $k$, this makes that $\frac{b(k)}{N^*_{ij}}$ is the density of flow per trip. The value $w(k)$ is the weight of the edge and $N_{ij}(k)$ is the number of trips between $i$ and $j$ such that $k$ belongs to its shortest path. The third factor is the number of intersections that they cross in that trip (which reduce the performance of the journey) similar to a measure of betweenness of intersections.
\begin{equation}
n=\frac{1}{N^*_{ij}}\sum_{\forall i}\sum_{\forall j}\frac{N_{ij}(k)(k(v)-2)}{\sigma_{ij}}
\end{equation}
where $k(v)$ is the degree of vertex $v\in V$. This factor is normalised by the number of outgoing trips from city to city in the system ($N^*_{ij}$) in order to have comparable measurements for all $\alpha$s.

\section{Further improving the model}
\subsection{Topography}
\begin{figure}
\centering
\includegraphics[width=1\textwidth]{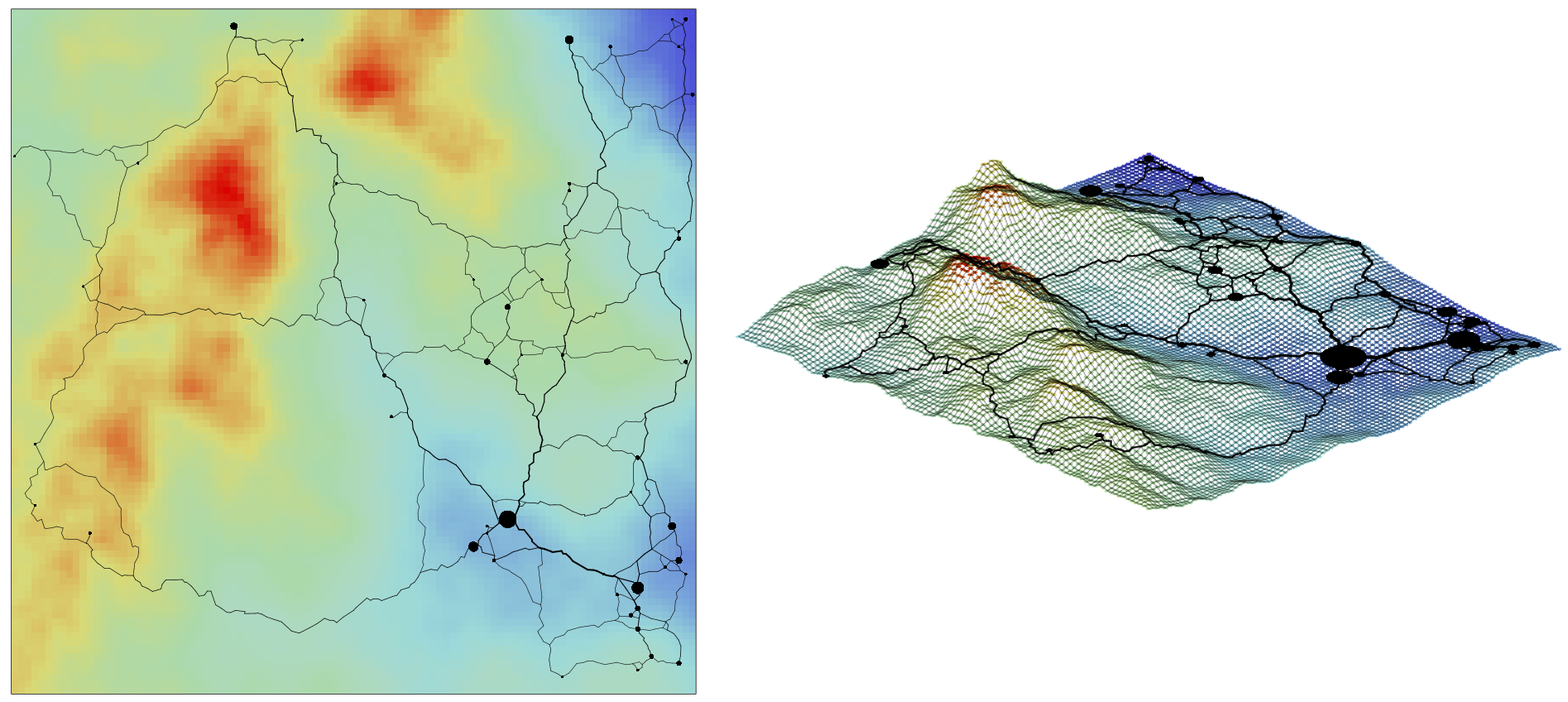}
\caption{Dealing with topography.
\label{fig::mountains}}
\end{figure}

In order to deal with the topography of the terrain, the only thing needed is to include multipliers for the speed. That is, for every edge in our delanuay (before calculating the roads), we calculate its absolute slope and use it to create a set of multipliers $M$ that will modify the final speeds of each segment.
Of course the larger the slope the smaller this multiplier should be, and they should never be 0. By doing this, the paths that have a smaller change in slope will be chosen as the shorter ones as shown in Fig.\ref{fig::mountains}. When the topography is very complex with abrupt changes, it is a good practice to increase the density of dummy nodes in those areas, so the algorithm can find the correct paths.

\subsection{Growth of cities}
\begin{figure}
\centering
\includegraphics[width=1\textwidth]{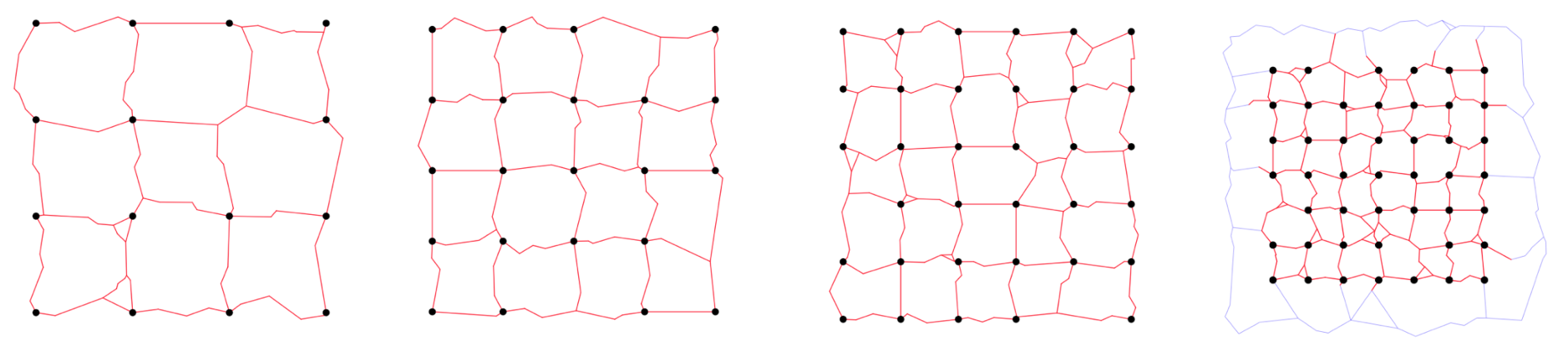}
\caption{Emergence of peripheral roads.
\label{fig::peripheralRoads}}
\end{figure}

If we were to grow one location at a time, and calculate the resulting set of roads, and at each time-step we calculate the boundaries of our city and set the multiplier of the edges within the city to match the reality (40km/h within cities 90km/h outside cities) at a given size of city peripheral roads would automatically appear, to connect the furthest away points of the city. This is shown in Fig.~\ref{fig::peripheralRoads}.
\subsection{Differentiating cities}
\begin{figure}
\centering
\includegraphics[width=1\textwidth]{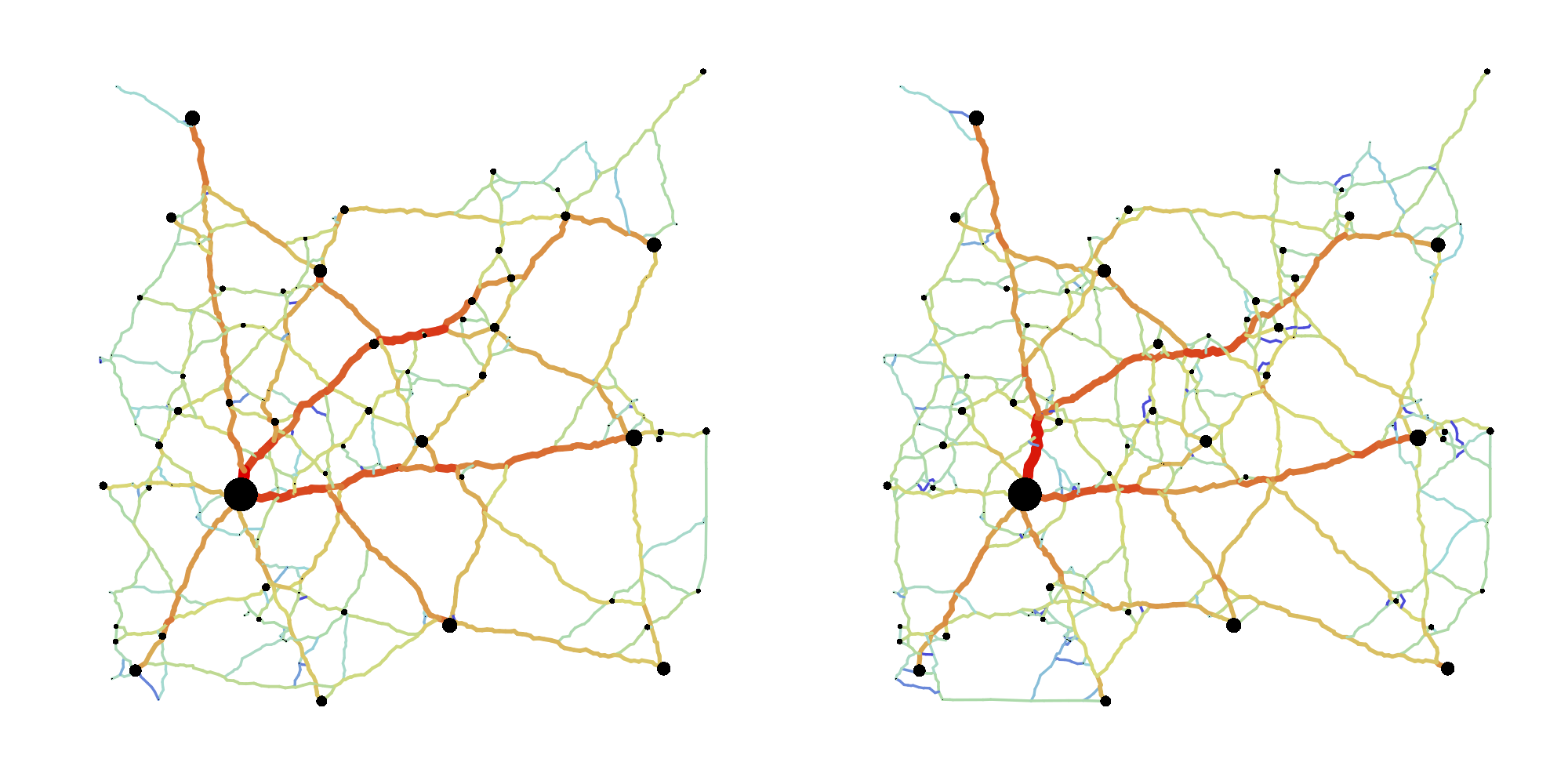}
\caption{Differentiating speeds of roads within cities and outside of them. a) Speeds are equal all over the set, no peripheral roads are created. b) considering different speeds within cities and outside of them with the consequent creation of peripheral roads.
\label{fig::dealingWithCities}}
\end{figure}

Similarly, when calculating a large region, if we are able to set multipliers for where city road segments are located, the algorithm would automatically calculate peripheral roads, to travel between cities, in order to avoid being slowed down by entering the city. We need to dynamically modify these weights such that when calculating the roads between cities $i,j$ the speeds of both cities are reset to normal, otherwise only a single entry point to the city will be produced. This is shown in Fig.~\ref{fig::dealingWithCities}

\subsection{Rivers}
\begin{figure}
\centering
\includegraphics[width=.5\textwidth]{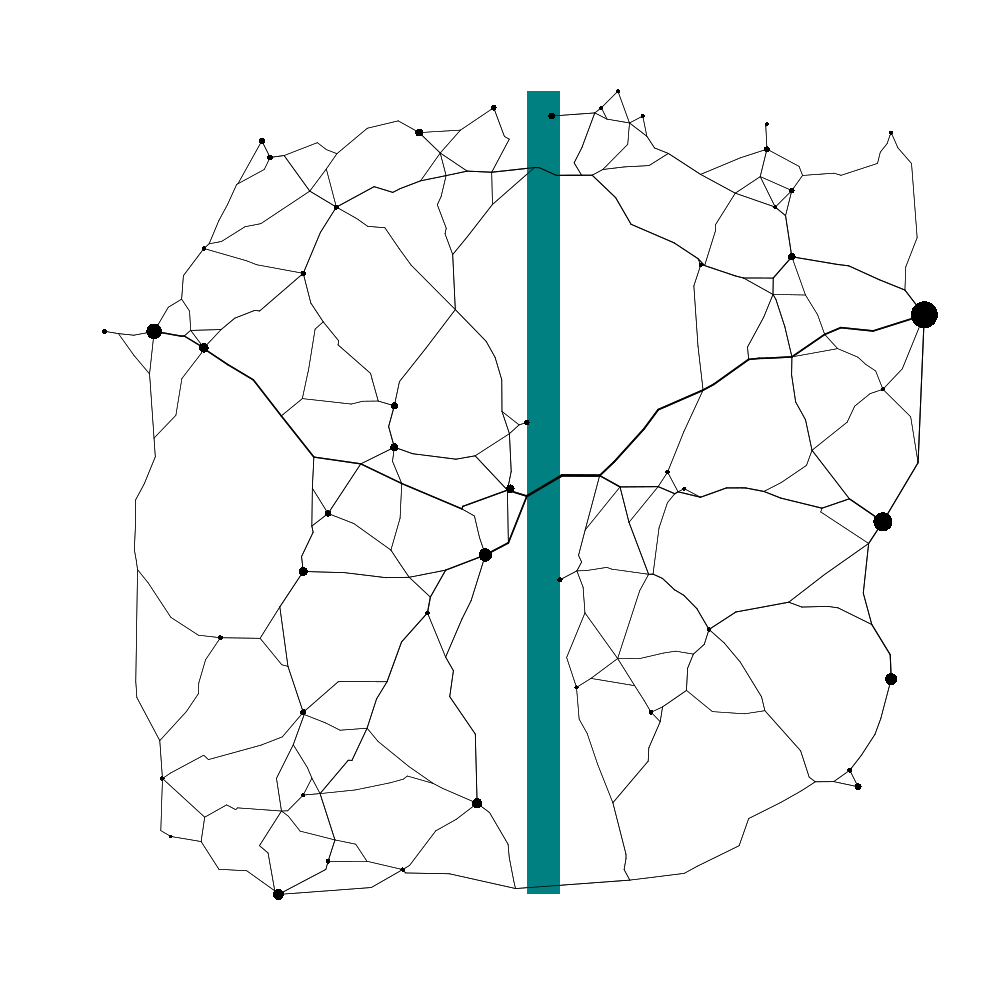}
\caption{Dealing with rivers, automated construction of bridges
\label{fig::rivers}}
\end{figure}

Rivers need to be handled differently. In this case, an edge of the delanuay that crosses the river should have a multiplier that decreases its speed depending on how difficult it is to build a bridge (width of the river), but the moment that the bridge is built, then its relative speed should return to normal. That is, when the algorithm constructs a (bridge) road segment, the multiplier of that edge should be set back to 1. An example is shown in Fig.\ref{fig::rivers}
\subsection{Airports and forbidden areas}
Dealing with any area that should not have roads (such as national parks, airports, etc.) is as simple as setting the multiplier of the speeds allowed to practically 0, which would make that any path growing through them would be discarded and other shorter paths would be chosen.

\subsection{Improving the final result}
\begin{figure}
\centering
\includegraphics[width=1\textwidth]{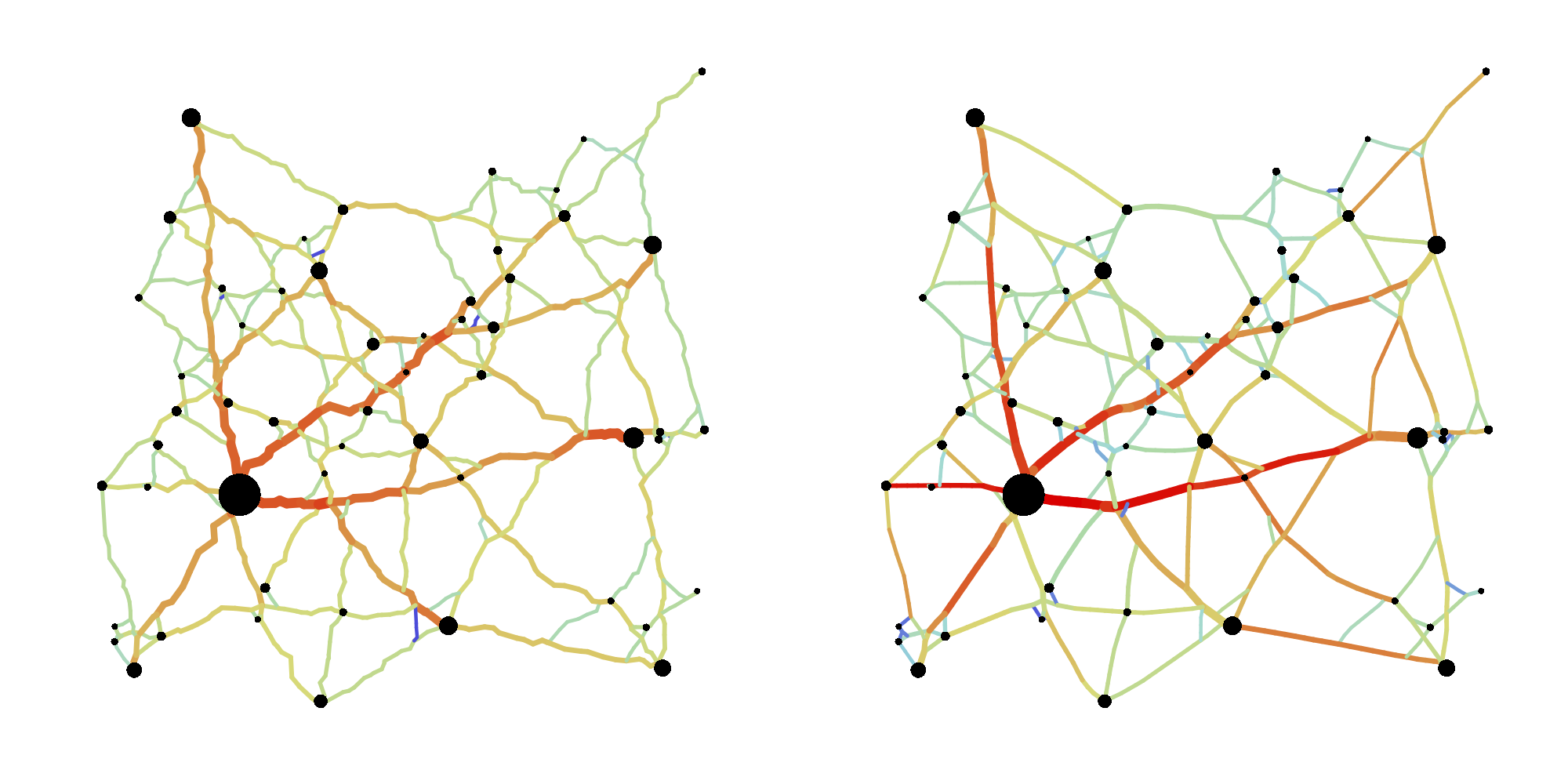}
\caption{Straightening the created set of roads.
\label{fig::straightenGraph}}
\end{figure}

Since the location of dummy nodes is produced randomly, and including a large number of nodes has a heavy effect on the speed of calculation, some artificial turns are included in the set of roads produced. In order to correct for this errors,
We have opted for calculating the centroid of each node (using the location of its neighbours) weighted by the edge local-weighted betweenness (the weighted average of the locations) and move the node a small step at a time towards that centroid. In order to avoid modifying too much the road networks, each node should only be allowed to move from its original location a small distance, to implement this, we weighted the movement towards the centroid by the inverse square of the distance to its original position. This is shown in Fig.~\ref{fig::straightenGraph}. Of course, much better solutions can be found, an important variation to include in the future is that nodes should move towards minimizing elevation change.

\end{document}